\documentclass[intlimits,twoside,a4paper]{article}

\usepackage{amsmath,amssymb}
\usepackage{graphicx}
\usepackage{wrapfig}

\usepackage[T2A]{fontenc}
\usepackage[cp1251]{inputenc}
\usepackage{multirow}
\usepackage{subfigure}

\usepackage[eqsecnum]{cmpj}
\newcommand{\EquationRef}[2]{#1 \eqref{#2}}
\newcommand{\FigureRef}[2]{#1 \ref{#2}}
\newcommand{\TableRef}[2]{#1 \ref{#2}}
\newcommand{\Angstr}{\AA$~$}
\newcommand{\OnLineRef}[1]{\cite{#1}}

\articletype{Regular article}

\title[3DRISM Bridge for Nanomaterials]%
{3DRISM Bridge Functional for the Aqueous Solutions of Carbon Nanomaterials%
}
\author{V.P. Sergiievskyi}
\address{ Max Planck Institute for Mathematics in the Sciences, 22 Inselstrasse, 04103 Leipzig, Germany
}
\begin{document}

\maketitle

\begin{abstract}

In the paper a bridge functional for the closure relation of the three-dimensional reference interaction model (3DRISM) equations is proposed. The effectiveness of the bridge for the aqueous solutions of the carbon nanomaterials is tested.
In the paper two classes of systems are investigated: (i) infinitely diluted aqueous solutions of the Lennard-Jones (LJ) spheres and (ii) infinitely diluted aqueous solution of the carbon nanotubes(CNT).
The bridge functional is fitted to the molecular simulation data.
It is shown that for all the investigated systems the bridge functional can be approximated by the exponential function which depends only on the solute/solvent size ratio.
It is shown that by using the proposed bridge functional is possible (i) to predict accurately the position of the first peak of the water oxygen density distribution function (DDF) around the solute, (ii) to improve the accuracy of the predictions of the first peak's height of the water oxygen DDF around the solute (iii),  to predict correctly the water hydrogen DDF behavior in the vicinity of the CNT. 
\pacs 61.20.Gy, 61.25.Em
\end{abstract}

\section{Introduction}

Integral equation theory of liquids (IETL) allows to calculate the thermodynamic parameters of a liquid and an amorphous matter by using the local microscopic structure which is determined by the density distribution functions (DDFs) \cite{Hansen2000}.
DDFs are straightforwardly connected to the total correlation functions which in turn can be calculated from the Ornstein-Zernike (OZ) equation ant the closure relation \cite{Ornstein1914,Hansen2000}.
For the molecular systems the correlation functions in a general case depend on the six variables. Solving the six-dimensional molecular OZ equation is still a challenging task.
In practice usually the approximations of the OZ equation are used such as the reference interaction site model (RISM) \cite{Chandler1972} and the three-dimensional reference interaction site model (3DRISM) \cite{Beglov1995}. For RISM and 3DRISM models the effective computational algorithms were developed\cite{Sergiievskyi2011, Fedorov2008b, Chuev2004, Kovalenko1999b,Sergiievskyi2011b}.
The RISM and 3DRISM equations allow to predict the properties of the liquid in a thermodynamic limit. In many applications they can be used as a substitution of the time-consuming molecular simulations. The 3DRISM was recently implemented in the Amber molecular modeling package and in the Amsterdam Density functional (ADF) quantum mechanics software\cite{Luchko2010,Gusarov2006}. This fact reflects the constantly increasing interest to this method in the chemical community. The 3DRISM can be used for investigating the structure of the molecular complexes \cite{Chuev2009, Yamazaki2010,Maruyama2011}, for determining  solvation effects on conformational equilibria \cite{Casanova2007,Lee2006}, for the accurate predicting of the solvation free energy \cite{Ratkova2010, Palmer2010, Palmer2010a, Palmer2011,Ratkova2011,Frolov2011,Sergiievskyi2011a}, and in many other applications \cite{Howard2011,Ten-no2010,Phongphanphanee2010,Kolombet2010a}. 
Although the RISM and 3DRISM equations predict qualitatively correctly the solvent structure. They often fail to reproduce the qualitative parameters such as the radial distribution functions (RDFs).
The main source of errors is not the model's approximations but the closure relation. 
The rigorous closure relation includes the so-called bridge functional which have in its analytical representation the infinite series of the integrals with the growing dimensionality\cite{Hansen2000}. This makes the bridge functional practically incomputable.
In practice the approximate closures are used. This causes the inaccuracies in the calculations. 
There was proposed a number of closure approximations. To name a few most popular: 
the Hyper Netted Chain (HNC) closure where the bridge functional is omitted, the Percus-Yevick (PY) closure for hard spheres \cite{Percus1958},  Martynov-Sarkisov closure \cite{Martynov1983}, the Verlet Modifiet closure \cite{Labik1991}, and others\cite{Martynov1992}.
The most of the functionals describe more or less accurately the simple liquids of hard spheres or LJ spheres. But even for the two-component solutions of LJ spheres of different size the effectiveness of the standard bridge functionals is questionable.

Another way to obtain the universal bridge functional is a parameterizing of the bridges which are obtained from the molecular simulations \cite{Du2000,Francova2011}.
One of the most simple but nevertheless effective method of the bridge functional parameterization is to approximate the bridge functional by the exponential function.
In the work \OnLineRef{Du2000} such kind of a parameterization was used to improve the predictions of the thermodynamical properties of organic compounds.
In the recent work \OnLineRef{Sergiievskyi2011b} it was shown that by using the exponential bridge functional it is possible to predict the solute-solvent RDFs of the LJ balls of different sizes dissolved in the LJ fluid. The parameters of the fluid were chosen to coincide with the water parameters at normal conditions.
This allows us to suggest that the exponential bridge functionals can be effective also for the aqueous solutions of LJ spheres. 
Moreover, in classical molecular dynamics (MD) simulations the carbon nanomaterials are described by the set of balls with LJ potential which allows us to expect the good performance of the exponential bridge functional also in this case.
In the current work the effectiveness of the exponential bridge functional for the aqueous solutions of carbon nanomaterials is tested.
To fit the parameters of the exponential bridge function the infinitely diluted aqueous solutions of the LJ spheres of different size  are used. 
The bridges are parameterized in oder to have the best coincidence of the solvent DDF which are obtained from the 3DRISM equations with the solvent DDF which are obtained from the MD simulations.

\section{Method}

\subsection{Exponential Bridge functional for the 3DRISM equations}

In the current paper the 3DRISM model is used in oder to describe the infinite diluted solutions of LJ balls and carbon nanomaterials.
The investigated systems contain one solute molecule surrounded by the solvent. 
The solute molecule is the three-dimensional object while the solvent molecules are represented in the RISM approximation as the set of interacting sites.

The 3DRISM equations are written in the following way:
\begin{equation}
\label{eq:3DRISM}
 h_{\alpha}(\mathbf{r}) = \sum_{\xi=1}^{N_{\rm site}} 
\int_{\mathbb{R}^3}
c_{\xi}(\mathbf{r}-\mathbf{r'})\chi_{\xi\alpha}(|\mathbf{r'}|) d\mathbf{r'}
\end{equation}
where $h_{\alpha}(\mathbf{r})$, $c_{\alpha}(\mathbf{r})$ are the total and the direct correlation functions  of the solvent site  $\alpha$ , $\chi_{\xi\alpha}(r)$ is the susceptibility function of the solvent sites $\xi$ and  $\alpha$.
The susceptibility function  $\chi_{\xi\alpha}(r)$ is defined in the following way: 
\begin{equation}
	\chi_{\alpha\xi}(r) = \omega_{\alpha\xi}(r) + \rho h_{\alpha\xi}(r)
\end{equation}
where $\rho$ is the solvent number density, $h_{\alpha\xi}(r)$ is the total correlation function of solvent sites $\xi$ and $\alpha$, $\omega_{\alpha\xi}$ is the solvent intramolecular function which defines the structure of the solvent molecule.
The solvent intramolecular function for the rigid molecules is defined in the following way:
\begin{equation}
 \omega_{\xi\alpha}(r) = 
	\delta_{\xi\alpha}\delta(r) + 
	(1-\delta_{\xi\alpha})
	\frac{\delta(r-r_{\xi\alpha})}{4\pi r_{\xi\alpha}^2}
\end{equation}
where $\delta_{\xi\alpha}$ is the Kronecker's delta, $\delta(r)$ is the Dirac delta function, $r_{\xi\alpha}$ is the distance between the  sites $\xi$ and $\alpha$ in the solvent molecule.

The 3DRISM equations \EquationRef{}{eq:3DRISM} are completed with the closure relation:
\begin{equation}
\label{eq:closure}
h_{\alpha}(\mathbf{r}) + 1 = 
e^{-\beta U_{\alpha}(\mathbf{r} + h_{\alpha}(\mathbf{r}) - c_{\alpha}(\mathbf{r}) + B_{\alpha}(\mathbf{r})}
\end{equation}
where $U_{\alpha}(r)$ is the potential of the interaction of the solute with the solvent site $\alpha$,
$B_{\alpha}(\mathbf{r})$ is the bridge functional of the solvent site $\alpha$.

As it was mentioned above the bridge functional  in general case is practically incomputable.
In the paper \OnLineRef{Sergiievskyi2011b} for the solutions of the LJ balls in the LJ fluid the following exponential bridge functional was used:
\begin{equation}
\label{eq:B_LJ}
  B(r) = -\exp(k(r-r_0) - C \sigma_{\rm solute} / \sigma_{\rm solvent})
\end{equation}
where $k$=30 nm${}^-1$,  C=1.18, $\sigma_{\rm solute}$ and $\sigma_{\rm solvent}$ are the $\sigma$ LJ parameters of solute and solvent atoms correspondingly, $r_0$ is the point where $U(r_0)=13.8 k_BT$. 
In the current work we use analogous bridge functional for the LJ spheres' aqueous solutions. 
Due to the spherical symmetry  the functions in the equations \EquationRef{}{eq:3DRISM},\EquationRef{}{eq:closure} depend only on the distance to the solute's center  $r=|\mathbf{r}|$.
In the current work we use the following bridge functionals:
\begin{equation}
\label{eq:B_alpha}
 B_{\alpha}(r) = -\exp(k(r-r_0^{\alpha}) - C \sigma / \sigma_0)
\end{equation}
where $\sigma_0$=0.316 nm is the oxygen $\sigma$ parameter in the SPC/E water model.

\subsection{Investigated systems}

In the current paper  two classes of  systems are investigated : (i) infinitely diluted aqueous solutions of the Lennard-Jones (LJ) spheres and (ii) infinitely diluted aqueous solution of the carbon nanotubes(CNT).
The $\sigma$ LJ parameters of the spheres was $\sigma=K \sigma_0$, where $K$=\{ 0.1, 0.2, 0.3, 0.4, 0.5, 0.6, 0.7, 0.8, 0.9, 1, 1.2, 1.3, 1.4, 1.5, 1.8, 2 \} , $\sigma_0=0.316$ nm.
For the investigated systems the RISM/3DRISM calculations as well as the MD simulations were performed.
For the MD simulations the program package  Gromacs 4.5.3 was used\cite{Hess2008}. 
LJ balls were put to the cubic box of size 4x4x4 nm filled with the water molecules. 
The SPC-E water model was used in the simulations \cite{Berendsen1987}.
The simulations were performed in the NVT ensemble with the temperature $T=300K$.
1 000 000 simulation steps of leap-frog integration method were performed. Each of the step corresponded to 1 femtosecond. 
For the RISM calculations the multi-grid algorithm described in the paper \OnLineRef{Sergiievskyi2011} was used.

The second class of systems are the aqueous solution of carbon nanotubes (CNT).
In that class investigate the single system was investigated: aqueous solution of the single-wall CNT.
The nanotube of chirality (7,0) was used which corresponds to the diameter of 0.55 nm.
The choice of the nanotube is explained by the fact that for the CNT of such chirality  MD simulation data is available in the literature (see supporting information for works \cite{Frolov2010,Frolov2011a}). 
To test the effectiveness of the bridge functionals the 3DRISM calculations for the same system were performed.
The LJ parameters of the OPLS force field was assigned to the carbon atoms of the CNT: $\sigma_C$=0.355 nm, $\epsilon_C$=0.07 kcal/mol.
For the calculations the multi-grid 3DRISM algorithm was used \cite{Sergiievskyi2011c}.
The calculations with the exponential bridges as well as without bridges (for comparison) were performed.
Because the 3DRISM equations without bridge cannot be solved numerically because of the divergence of the algorithm for the system without bridge the Kovalenko-Hirata closure was used \cite{Kovalenko1999}. 
This closure is the standard closure for the 3DRISM calculations \cite{Kovalenko1999}. 
In the 3DRISM calculations the Cartesian grid is used. It is characterized by two values: the buffer, which is the minimal distance from the solute atoms to the boundaries of the grid and the spacing which is the minimal distance between the grid points.
For the 3DRISM calculations the grid with the buffer 1.5 nm and spacing 0.02 nm was used.
In the RISM and 3DRISM calculations the MSPC/E water model was used \cite{Lue1992}. The water hydrogen had the following LJ parameters:  $\sigma_H$=0.8\Angstr, $\varepsilon_H$=0.046 kcal/mol.

\subsection{Parameterization of the bridge functional}

In the work \cite{Sergiievskyi2011b} the optimal values for $k$ and $C$ parameters in \EquationRef{the equation}{eq:B_LJ} were determined.
Because of the similarity in the solvent LJ parameters in the current paper the same values as in the work  \cite{Sergiievskyi2011b} are used:  $k$=30 nm${}^-1$, C=1.18.
The $r_0^{\alpha}$  parameter depends on the shape of the potential and LJ parameters of the atom pair, and thus is not universal.
In the current work the optimal values of $r_0^{\alpha}$ are determined and connected to the universal parameters, such as the positions of minimum of the potentials $U_{O}^{\sigma}(r)$, $U_{H}^{\sigma}(r)$ which describe the interaction of the water oxygen and water hydrogen with the solute.
For the parameterization is more convenient to use not the functions  $U_{O}^{\sigma}(r)$, $U_{H}^{\sigma}(r)$ but dimensionless functions $\exp(-\beta U_{O}^{\sigma}(r))$, $\exp(-\beta U_{H}^{\sigma}(r))$. They are the first approximations of the RDF functions (if one considers only the direct atom-atom interactions neglecting the influence of the solvent).
 The maximums' positions of the functions $\exp(-\beta U_{O}^{\sigma}(r))$ and  $\exp(-\beta U_{H}^{\sigma}(r))$ in the current work are denoted as $r^{\rm exp}_{\sigma O}$ and $ r^{\rm exp}_{\sigma H} $  correspondingly. 
In the current work the correlation of the parameters $r_0^{O}$, $r_0^{H}$  with the $ r^{\rm exp}_{\sigma O} $, $ r^{\rm exp}_{\sigma H} $ for different sizes of the solute is investigated.

It is supposed that the following relations hold:
\begin{equation}
\label{eq:r_0}
	r_0^{O} = A r^{\rm exp}_{\sigma O}, r_0^{H} = A r^{\rm exp}_{\sigma H}
\end{equation}
where the constant $A$ is the same for the water oxygen and water hydrogen, but can be different for different sizes of the solvent LJ ball: $A=A(\sigma_{\rm solute}/\sigma_{\rm solvent})$

For each solute size  $\sigma_{\rm solute}$ the values of $A$ from $A=0.5$ to $A=1$ are tested.
For each of the $A$ values the quality of the local solvent structure prediction is estimated.
The center of mass of the water molecule nearly coincides with the oxygen atom which allows to estimate the local solvent structure prediction using the water oxygen RDF functions.
The position and the height of the first peak of the water oxygen RDF as the numerical characteristic of the prediction of the RDF were used. These two parameters are very important for the thermodynamics of solution.
It is reasonable to suggest that among these two characteristics the peak's position is more important.
It describes the structure of the solvation shell while the peak's height is only the quantitative characteristic of the solvation shell's density.
Thus,  the first peak's position was used in order to find the best $r_0^{O}$, $r_0^{H}$ values. After that it was tested how good the obtained bridge functional predicts the first peak's height of the RDF.

\section{Results}

RISM calculations for the infinitely diluted aqueous solutions of LJ sphere were performed.
In the calculations the bridge functionals \EquationRef{}{eq:B_LJ} were used. The $r_0^{O}$, $r_0^{H}$ were chosen using \EquationRef{the formula}{eq:r_0} and the parameter $A$ was varying in the range from $A=0.5$ to $A=1$.

\begin{figure}
\center
\includegraphics[width=0.8\textwidth]{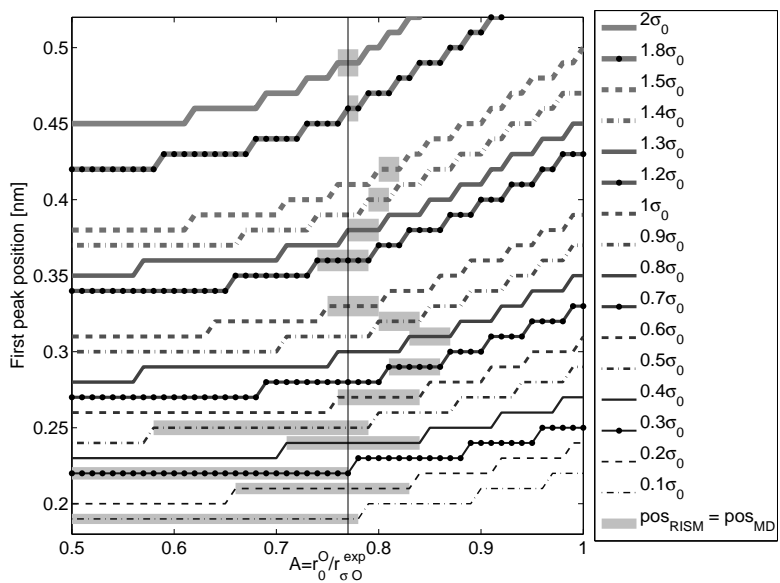}
\caption{
\label{fig:FirstPeakPos} Dependence of the first peak position of the water oxygen RDF calculated from the RISM equations with the exponential bridge  \EquationRef{}{eq:B_alpha} on the ratio $A=r_0^{O}/r^{\rm exp}_{\sigma O}$.
The zones where the first peak position of the RDF obtained from the RISM coincides with the first peak position obtained from the MD simulations are marked with the gray color.
}
\end{figure}

In the \FigureRef{Figure}{fig:FirstPeakPos} the dependence of the first peak position of the water oxygen  RDF on the ratio $A=r_0^{O}/r^{\rm exp}_{\sigma O}$ is presented.
The zones where the first peak position of the RDF obtained from the RISM coincides with the first peak position obtained from the MD simulations are marked with the gray color.
As it can be seen, for the LJ spheres of small size the first peak of the water oxygen  RDF weakly depends on $A$ while for larger spheres the optimal value of $A$ slightly fluctuates around the mean value.
This allows to suggest that the approximation $A=const$ can quite good fit to the MD-simulation's results.
To determine the optimal  value of $A$ for each value of $A$ the total error as a sum of the errors for different  solute sizes was found:
\begin{equation}
\label{eq:error}
  error(A) = \sum_{\sigma} | r_{\sigma}^{MD} - r_{\sigma}^{RISM}(A) |
\end{equation}
where $r_{\sigma}^{MD}$ is the first peak position of water oxygen RDF obtained from the MD simulations, $r_{\sigma}^{RISM}(A)$ is the first peak position of water oxygen RDF obtained from the RISM equations \EquationRef{}{eq:3DRISM} and the closure relation \EquationRef{}{eq:closure}, where the bridge functionals are calculated by using the formula \EquationRef{}{eq:B_alpha}, and the parameters $r_0^{O}(A)$, $r_0^{H}(A)$ are connected to the positions of the first maximum of the functions $\exp(-\beta U_{O}^{\sigma}(r))$, $\exp(-\beta U_{H}^{\sigma}(r))$ by the expressions \EquationRef{}{eq:r_0}.

\begin{figure}
\center
\includegraphics[width=0.6\textwidth]{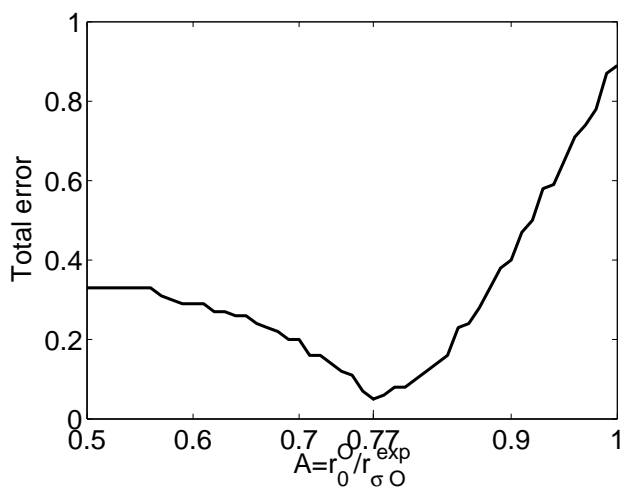}
\caption{\label{fig:error} 
Dependence of the total error of the first peak's position of the water oxygen RDF prediction \EquationRef{}{eq:error}  on the ratio $A=r_0^{O}/r^{\rm exp}_{\sigma O}$.
}
\end{figure}

In the \FigureRef{Figure}{fig:error} the dependence of the total error of the first peak position prediction \EquationRef{}{eq:error}  on the parameter $A$ is presented.
The minimal error corresponds to the value $A=0.77$.
So, the optimal universal bridge functional is the following: 
\begin{equation}
\label{eq:B_alpha_best}
 B_{\sigma\alpha}(r) = -\exp(k(r- A\cdot r^{\rm exp}_{\sigma \alpha}) - C \sigma / \sigma_0)
\end{equation}
where $A=0.77$, $r^{\rm exp}_{\sigma \alpha}$ is the position of maximum of the function $\exp(-\beta U_{\alpha}^{\sigma}(r))$.

\begin{figure}
\includegraphics[width=0.8\textwidth]{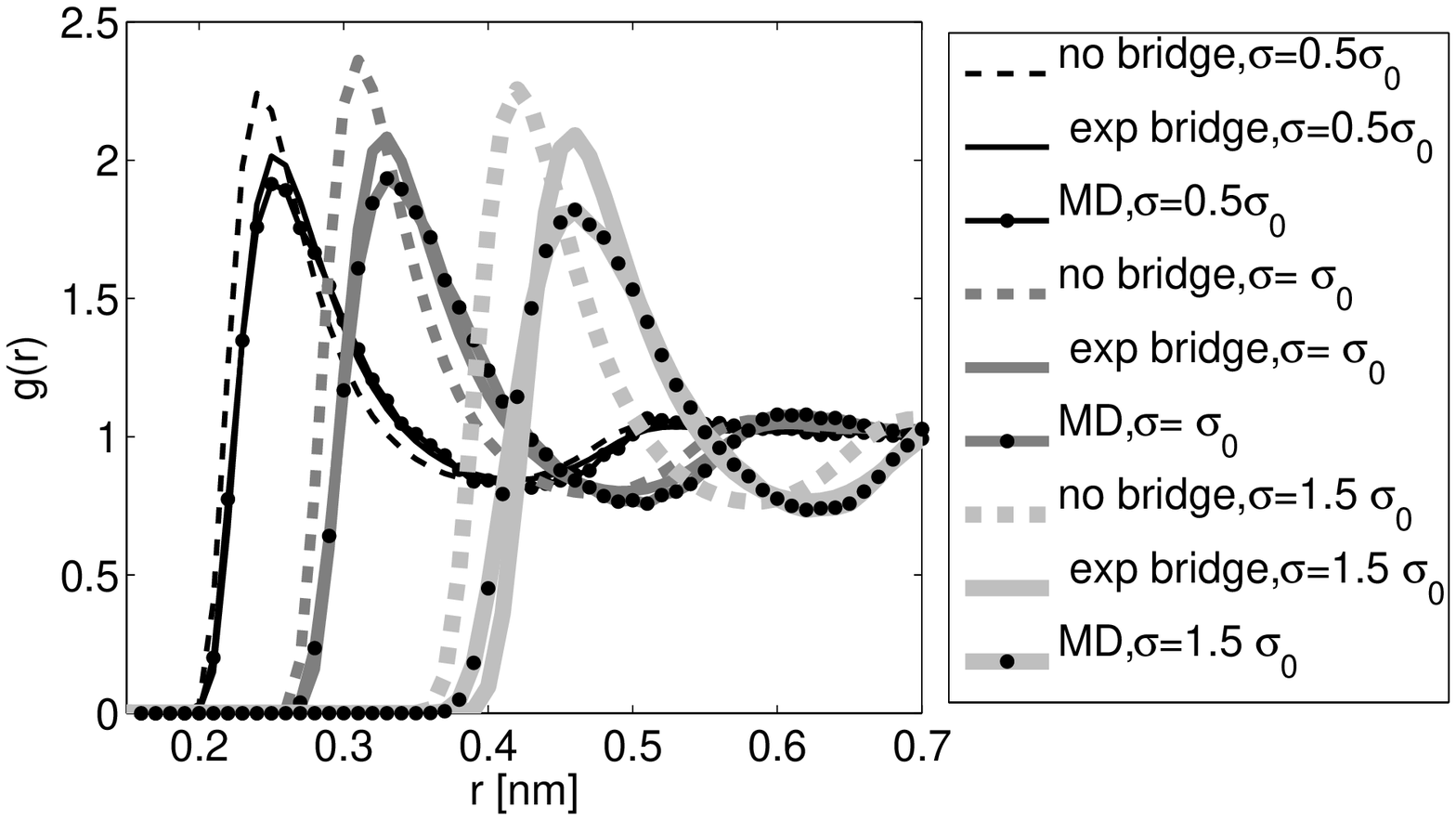}
\caption{\label{fig:RISMRDFs} 
The the water oxygen RDFs obtained from the RISM equations with bridge, from the RISM equations without bridge and from the MD-simulations.
Three groups of RDFs for spherical LJ solutes of different sizes are presented: $\sigma_{\rm solute}$=0.5$\sigma_0$,  $\sigma_{\rm solute}$=$\sigma_0$, $\sigma_{\rm solute}$=$1.5\sigma_0$, where  $\sigma_0$ = 0.316 nm is the $\sigma$ LJ parameter of the water oxygen in the SPC/E water model.
}
\end{figure}

\begin{table}
\begin{tabular}{ | c | c |  c |  c |  c |  c | }
\hline
\multirow{2}{*}{$\sigma/\sigma_{0}$}&
\multirow{2}{*}{\parbox[c]{3cm}{Peak height in MD} }& 
\multicolumn{2}{|c|}{No bridge } \vspace{1pt}& 
\multicolumn{2}{|c|}{With the exponential bridge } \vspace{1pt} \\ \cline{3-6}
& & \vspace{1pt} \parbox[c]{2cm}{Peak height}  & \vspace{1pt} \parbox[c]{2cm}{Difference to MD} &
\vspace{1pt} \parbox[c]{2cm}{Peak height}  & \vspace{1pt} \parbox[c]{2cm}{Difference to MD} \\
\hline
0.1&    1.745&    1.824&    0.079&    1.645&    -0.101\\
0.2&    1.799&    1.935&    0.135&    1.746&    -0.053\\
0.3&    1.860&    2.007&    0.148&    1.842&    -0.017\\
0.4&    1.916&    2.144&    0.229&    1.947&    0.031\\
0.5&    1.914&    2.243&    0.328&    2.015&    0.101\\
0.6&    1.926&    2.296&    0.370&    2.064&    0.138\\
0.7&    1.977&    2.358&    0.381&    2.085&    0.108\\
0.8&    1.944&    2.366&    0.421&    2.076&    0.132\\
0.9&    1.904&    2.348&    0.444&    2.089&    0.185\\
1.0&    1.934&    2.363&    0.428&    2.085&    0.150\\
1.2&    1.862&    2.343&    0.480&    2.093&    0.231\\
1.3&    1.873&    2.338&    0.465&    2.087&    0.214\\
1.4&    1.888&    2.311&    0.423&    2.094&    0.206\\
1.5&    1.866&    2.318&    0.452&    2.091&    0.225\\
1.8&    1.820&    2.256&    0.436&    2.091&    0.270\\
2.0&    1.794&    2.208&    0.414&    2.100&    0.306\\
\hline
Average &        1.876&    2.229&    0.352&    2.009&    0.133\\
\hline
\end{tabular}
\caption{\label{tab:FirstPeakHeight} 
Comparison of the first peak's height of the oxygen water RDFs obtained  by the three methods:  the RISM simulations with bridge,  the RISM simulations without bridge and the MD simulations.
}
\end{table}

\begin{table}
\begin{tabular}{ | c | c |  c |  c |  c |  c | }
\hline
\multirow{2}{*}{$\sigma/\sigma_{0}$}&
\multirow{2}{*}{\parbox[c]{3cm}{Peak position in MD} }& 
\multicolumn{2}{|c|}{No bridge} \vspace{1pt}& 
\multicolumn{2}{|c|}{With exponential bridge} \vspace{1pt} \\ \cline{3-6}
& & \vspace{1pt} \parbox[c]{2cm}{Peak position [nm]}  & \vspace{1pt} \parbox[c]{2cm}{Difference to MD [nm]} &
\vspace{1pt} \parbox[c]{2cm}{Peak position [nm]}  & \vspace{1pt} \parbox[c]{2cm}{Difference to MD [nm]} \\
\hline
0.1&    0.19&    0.19&    0.00&    0.19&    0.00\\
0.2&    0.21&    0.20&    -0.01&    0.21&    0.00\\
0.3&    0.22&    0.22&    0.00&    0.22&    0.00\\
0.4&    0.24&    0.23&    -0.01&    0.24&    0.00\\
0.5&    0.25&    0.24&    -0.01&    0.25&    0.00\\
0.6&    0.27&    0.26&    -0.01&    0.27&    0.00\\
0.7&    0.29&    0.27&    -0.02&    0.28&    -0.01\\
0.8&    0.31&    0.28&    -0.03&    0.30&    -0.01\\
0.9&    0.32&    0.30&    -0.02&    0.31&    -0.01\\
1.0&    0.33&    0.31&    -0.02&    0.33&    0.00\\
1.2&    0.36&    0.34&    -0.02&    0.36&    0.00\\
1.3&    0.38&    0.35&    -0.03&    0.38&    0.00\\
1.4&    0.40&    0.37&    -0.03&    0.39&    -0.01\\
1.5&    0.42&    0.38&    -0.04&    0.41&    -0.01\\
1.8&    0.46&    0.42&    -0.04&    0.46&    0.00\\
2.0&    0.49&    0.45&    -0.04&    0.49&    0.00\\
Average &    &    &    -0.021&    &    -0.003\\
\hline
\end{tabular}
\caption{\label{tab:FirstPeakPos}
Comparison of the first peak's position of the oxygen water RDFs obtained  by the three methods:  the RISM simulations with bridge, the RISM simulations without bridge, and the MD simulations.
}
\end{table}

To test the bridge functional \EquationRef{}{eq:B_alpha_best}   the RISM calculations of the infinitely diluted aqueous solutions of LJ spheres of different size with the bridge functional \EquationRef{}{eq:B_alpha_best}  and without the bridge ($B_{\sigma\alpha}(r) \equiv 0$) were performed. The RDFs were compared to the RDFs obtained from the MD simulations.
In \TableRef{Table}{tab:FirstPeakHeight} the comparison of the first peak's height is presented. 
In \TableRef{Table}{tab:FirstPeakPos} the comparison of the first peak's position is presented.
For the three selected sizes of the solute LJ spheres the oxygen water RDFs obtained from the RISM equations (with and without bridge) and from the MD simulations are presented in \FigureRef{Figure}{fig:RISMRDFs}.
It can be seen that by using the bridge functional \EquationRef{}{eq:B_alpha_best} it is possible to predict the   water oxygen RDF's first peak's position  with the average accuracy of 0.003 nm (0.03\Angstr). This is 3 times less than the resolution of the grid used for discretization the RDFs (0.1\Angstr).
Without using the bridge functional the average error of the peak's position prediction is 0.021 nm (0.21\Angstr) which is essential e.g. for the solvation free energy calculations.
Introducing the bridge functional \EquationRef{}{eq:B_alpha_best} also improves the predictions of the first peak's height of the oxygen water RDF. Without the bridge functional the peak's height is overestimated in average by 0.352 units (18.76\%) while with the bridge functional the average error is only 0.133 units (7.1\%).

In conclusion: introducing the bridge functional improves the accuracy of RISM calculations for aqueous solutions of the LJ spheres of different size.
This also allows to suggest the effectiveness of the proposed bridge functional for the aqueous solutions of the carbon nanomaterials.
To check this assumption the 3DRISM calculations for the aqueous CNT solution were performed with the proposed bridge functional and without it (using the Kovalenko-Hirata closure \cite{Kovalenko1999}).
The bridge functional for the nanotube was calculated as the superposition of the bridge functionals of the atoms of the nanotube:
\begin{equation}
	B_{\alpha}(\mathbf{r}) = \sum_s B_{\sigma \alpha} (|\mathbf{r} - \mathbf{r}_s|)
\end{equation}
where $\mathbf{r}_s$ is the position of the atom $s$ of the CNT,  $B_{\sigma \alpha} (r)$ is defined by the formula \EquationRef{}{eq:B_alpha_best} where $\sigma=\sigma_C$=0.355 nm, and the sum is calculated over all the atoms of the nanotube.
I calculated the mean number densities of the water oxygen and water hydrogen at the different distances from the nanotube axis.
Comparison of the RISM calculations with the results of the MD-simulations which were done in the work \OnLineRef{Frolov2010} are presented in the \FigureRef{Figure}{fig:CNT}.

\begin{figure}
\centering
\subfigure[
Water oxygen atoms' density distribution around the nanotube's axis
]{\includegraphics[width=0.45\textwidth]{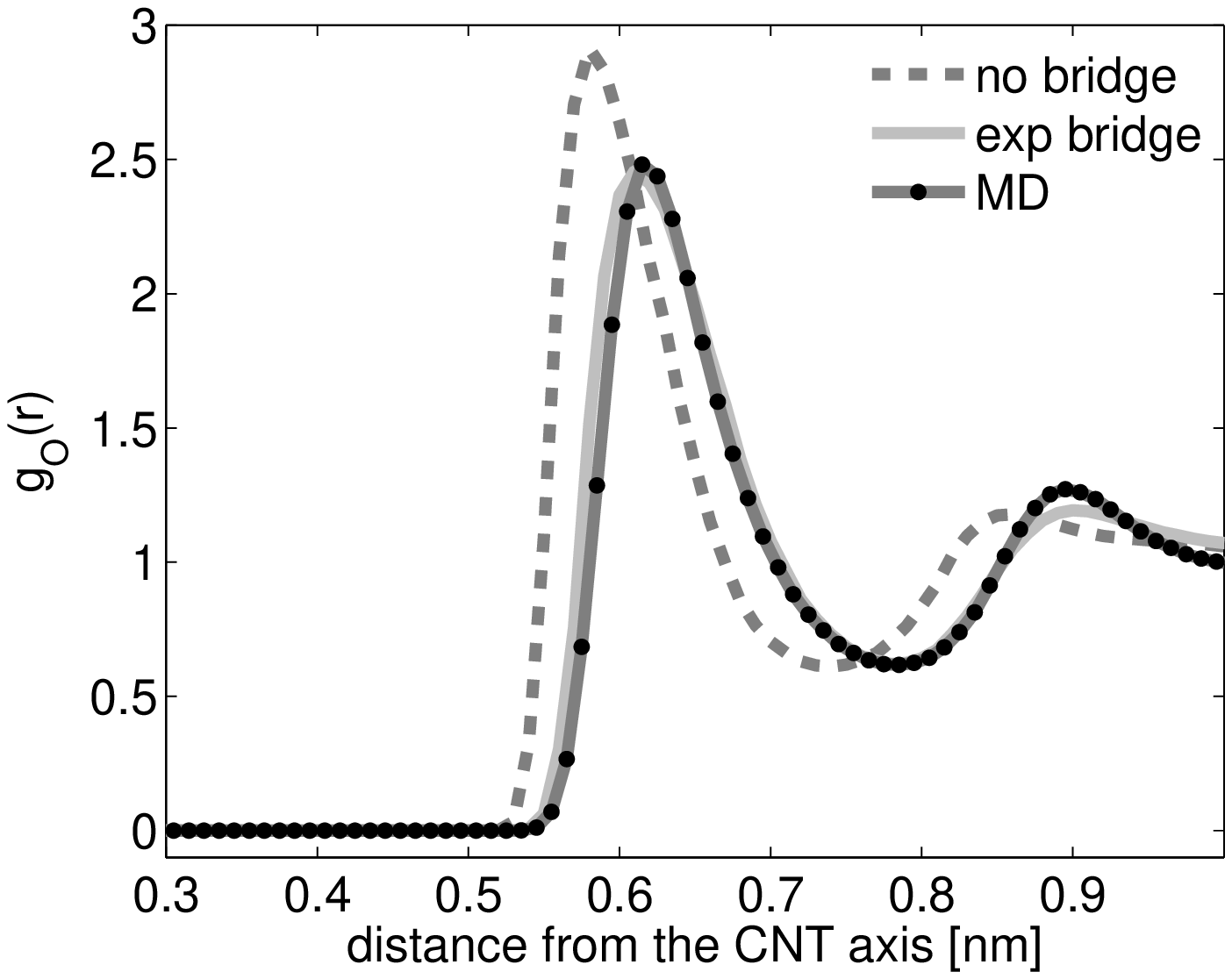} }
\subfigure[
Water hydrogen atoms' density distribution around the nanotube's axis
]{\includegraphics[width=0.45\textwidth]{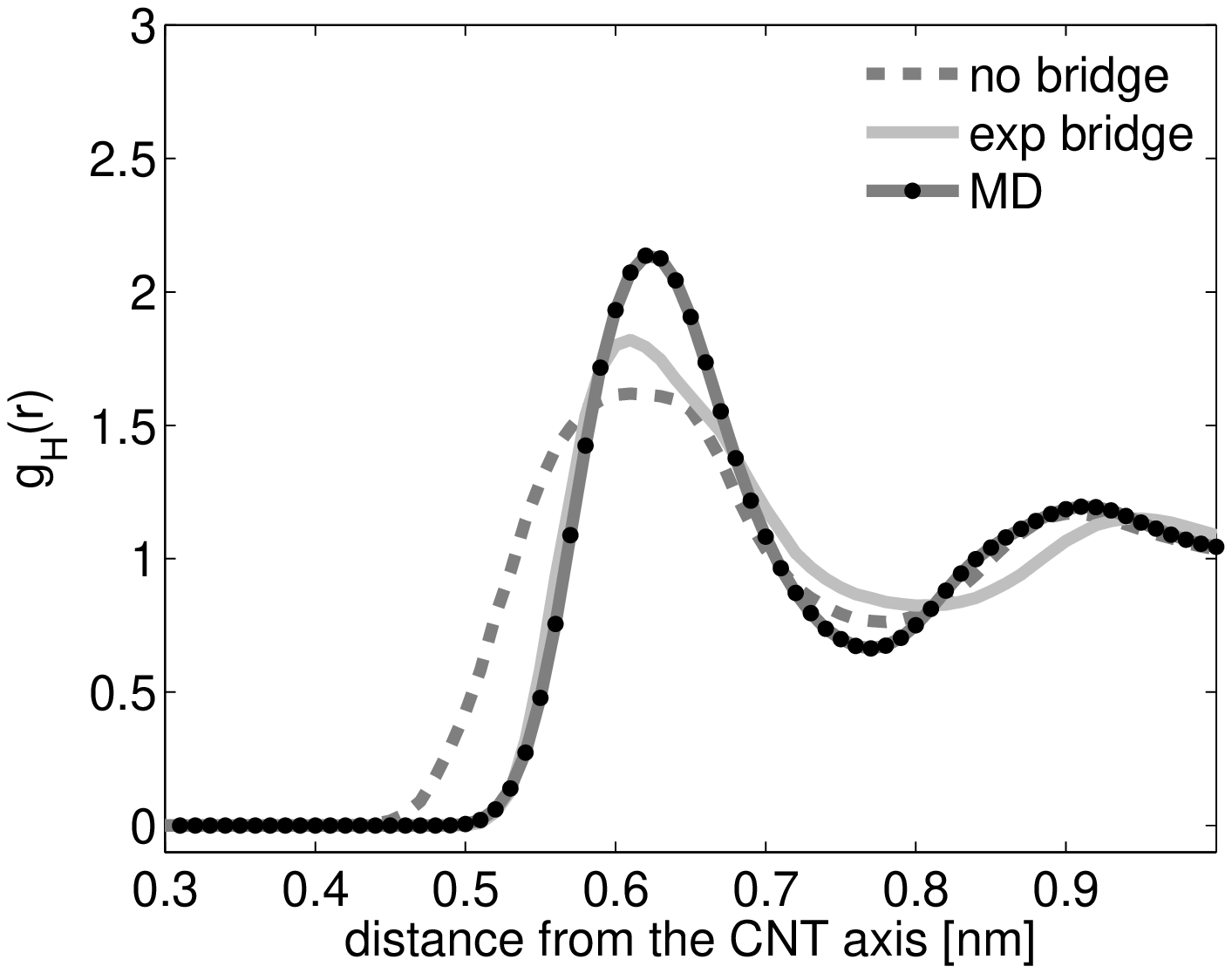} }
\caption{\label{fig:CNT} 
Comparison of the RISM calculations (with the exponential bridge and without the bridge)  with the results of the MD-simulations for the CNT aqueous solution
}
\end{figure}

As it can be seen, introducing the bridge functional allows to predict almost accurately the distribution of the water oxygen around the axis of CNT.
The center of mass of the water molecule nearly  coincide with the center of oxygen atom. That means that the 3DRISM with the proposed bridge functional can also correctly predict the water density around the carbon nanotube.
The prediction of the distribution of the hydrogen atoms is not perfect. 
This can be explained by the fact that the 3DRISM  describes the structure of the solvent molecule not enough accurate. 
However, it is necessary to notice that in the vicinity of the nanotube (at the distances <0.6 nm from the nanotube's axis) the water hydrogen distribution functions obtained from the 3DRISM calculations nearly coincide with the water hydrogen distribution functions obtained from the MD simulations.
This fact allows to conclude that the bridge functional allows to predict correctly the water hydrogen interaction  with the CNT.

\section{Conclusions}

In the current work the exponential bridge functional in the closure of the 3DRISM equation was parameterized and tested.
Two classes of systems were investigated: the infinitely diluted aqueous solution of the LJ spheres and the infinitely diluted aqueous solution of the carbon nanotubes.
The following $\sigma$ parameters of the LJ spheres were used: $\sigma= K \sigma_0$, where $K=$ \{ 0.1, 0.2, 0.3, 0.4, 0.5, 0.6, 0.7, 0.8, 0.9, 1, 1.2, 1.3, 1.4, 1.5, 1.8, 2 \}, $\sigma_0=0.316$ nm.
For all of the systems the MD simulations and the RISM calculations with the bridge functional \EquationRef{}{eq:B_alpha} and different values of $r_0^{\alpha}$ were performed. 
It was shown that introducing the bridge functional can essentially improve the prediction of the the water oxygen RDF's first peak's position.
The dependency of the optimal values of $r_0^{\alpha}$ on the size of the LJ solute was investigated. 
It was shown that  $r_0^{\alpha}$  is related to the position of the maximum of the function $\exp(-\beta U_{\sigma\alpha}(r))$, where $U_{\sigma\alpha}(r)$ is the interaction potential of the LJ sphere of size $\sigma$ with the water's atom $\alpha$.
The optimal value of constant $A$ in the relation \EquationRef{}{eq:r_0} was determined.
As a result the universal functional  \eqref{eq:B_alpha_best} was proposed.
It was shown that by using the bridge functional  \EquationRef{}{eq:B_alpha_best} it is possible to predict the the water oxygen RDF's first peak's position  with the accuracy of 0.003 nm (0.03\Angstr). This is 3 times less than the resolution of the grid which was used for discretization of the RDFs.
The RISM equations without the bridge give a systematic error of the first peak's position prediction of 0.021 nm (0.21\Angstr).
It was shown that using the bridge functional \EquationRef{}{eq:B_alpha_best} also improves the predictions of the first peak's height of the water oxygen RDFs.
Without the bridge functional the first peak's height is overestimated in average by 18.76\% while with the bridge functional the average error is only 7.1\%.
The effectiveness of the proposed bridge functional was also tested on the more complex system: the aqueous carbon nanotube solution.
For the investigations the carbon nanotube of the chirality (7,0) (diameter 0.55 nm) was used.
The water oxygen and water hydrogen density distributions at the different distances from the carbon nanotube were calculated.
It was shown that the bridge functional allows almost accurately predict the water oxygen atom's distribution around the CNT (and thus the water molecules' distribution, because the center of mass of the water molecule is nearly coincides with the center of the oxygen atom).
It was shown that the bridge functional allows to obtain the correct water hydrogen distribution in the vicinity of the CNT. 

\section*{Acknowledgements}

I would like to acknowledge my supervisor Maxim V. Fedorov for the support of the research.
I also would like to acknowledge Andrey I. Frolov for the consultations and help in performing the MD simulations. 
I would like to acknowledge the Max Planck Institute for Mathematics in the Sciences for the financial and technical support of the research.
The work was performed in the  framework of the BioSol program ( REA Research Executive Agency, Grant No. 247500 FP7-PEOPLE-2009-IRSES).

%
%% If you have problems with typesetting in ukrainian uncomment lines below.
%
%  \lastpage
%  \end{document}

\ukrainianpart

\title{Місткова функція тривимірної моделі взаємодіючих центрів (3DRISM) для водних розчинів вуглецевих наноматеріалів}
\author{В.П. Сергієвський}
\address{Max Planck Institute for Mathematics in the Sciences, 22 Inselstrasse, 04103 Leipzig, Germany}
%
%% якщо автор є один або автори є з однієї установи:
%
%  \author{1й Автор, 2й Автор, \ldots}
%  \address{Інститут\ldots}
%
%%

\makeukrtitle

\begin{abstract}
\tolerance=3000%
У статті запропановано новий містковий функціонал для рівнянь замкнення тривимірної моделі взаємодіючих центрів (3DRISM).
Тестується еффективність місткового функціоналу для водних розчинів вуглецевих наноматеріалів.
У статті розглядаються два класа систем: (1) нескінченно розбавлені водні розчини Леннард-Джонсовських сфер (2) нескінченно розбавлені водні розчини вуглецевих нанотрубок.
Містковий функціонал параметеризується виходячи із данних симуляцій. 
Показано, що для усіх досліджених систем містковий функціонал може бути наближений експоненціальною функцією, що залежить тільки від співвідношення розмірів розчиненої речовини та розчинника.
Показано, що використовуючи запропонований містковий функціонал є можливим: (1) точно завбачити положення першого піку функцій розподілення щильності (ФРЩ) кисню води навколо розчиненої речовини (2) покращити точність завбачень висоти першого піку ФРЩ кисню води навколо розчиненої речовини (3) правильно завбачити поведінку ФРЩ водню води поблизу нанотрубки.
\keywords 3DRISM, містковий функціонал, вуглецевi нанотрубки, функції розподілення щильності

\end{abstract}

\end{document}